\documentstyle[12pt]{article}
\begin{document}
\parskip 10pt plus 1pt
\title{VAN DER WAERDEN'S COLOURING THEOREM AND CLASSICAL SPIN
SYSTEMS}
\author{
{\it  Ranjan Chaudhury, Debashis Gangopadhyay  and  Samir K. Paul}
\\{S.N.Bose National Centre For Basic Sciences}\\
{JD Block, Sector-III, Salt Lake, Calcutta-700091, INDIA}\\
}
\baselineskip=20pt
\date{}
\maketitle

\begin{abstract}
We find a non-invertible matrix representation for  Van der Waerden's 
colouring theorem for two distinct colours in a one dimensional
periodic lattice.Using this,an infinite one dimensional antiferromagnetic
Ising system is mapped to a pseudo-ferromagnetic one, thereby relating the 
couplings.All this is reminiscent of renormalisation group.
PACS NO: 75.10 Hk
\end{abstract}
\vskip 2in

\newpage
One of the many equivalent versions of Van der Waerden's Colouring Theorem in 
combinatoric number theory is $^{1)}$ :

Let $X$ be a finite subset of $N^{d}$. For each m-colouring of $N^{d}$ there
exists a positive integer $\alpha$ and a point $\beta$ in $N^{d}$ such
that the set $\alpha X + \beta$ is monochromatic. Moreover, the
number $\alpha$ and the coordinates of the point $\beta$ are
bounded by a function that depends only on $X$ and $m$ (and not
on the particular colouring used).

Here we find a matrix representation of Van der Waerden's (VDW) colouring 
theorem for two distinct colours in a one dimensional periodic lattice.The
representation turns out to be non-invertible.The two colours are identified 
as spin "up" and spin "down", and an explicit mapping of an infinite one 
dimensional antiferromagnetic Ising $^{2)}$  system to a pseudo-ferromagnetic 
one is obtained.Consider the antiferromagnetic spin ordering in an 
infinite one dimensional lattice.

We consider $N^{1}$ and our subset $X$ consists of
any two points of the one dimensional lattice such that the
first point has ``up'' spin and the second a ``down'' spin. If
$k$ denotes the point with an ``up'' spin then it necessarily
follows that the ``down'' spin will occur at a spacing $k + (2m
+ 1)b, m = 0,1,2,......$ For example, if $m = 0$ then there are
two nearest neighbour points with separation $b$. The set can be
anywhere on the lattice. The coupling between any two spins is
taken to be some function of their separation.

To simulate the infinite lattice we use periodic boundary
conditions i.e. R..B..R..B..etc. ``Up''
spins are labelled by $R$ (red) and the ``down'' spins by $B$ (blue).

VDW theorem in the context of the above spin chain may be
quantified as follows :
$$\left(\begin{array}{l}
R_{\alpha k+\beta}\\
R_{\alpha[k+(2m+1)b]+\beta}\\
\alpha k + \beta\\
\alpha[k+(2m+1)b]+\beta\end{array}\right) = 
\left(\begin{array}{cccc}
V_{11} & V_{12} & V_{13} & V_{14}\\
V_{21} & V_{22} & V_{23} & V_{24}\\
V_{31} & V_{32} & V_{33} & V_{34}\\
V_{41} & V_{42} & V_{43} & V_{44}\end{array}\right) 
\left(\begin{array}{l}
R_{k}\\
B_{k+(2m+1)b}\\
k\\
k+(2m+1)b \end{array}\right) \eqno(1)$$
where for generality we have assumed the lattice spacing in the subset $X$ to be
$(2m + 1)b$\enskip   ($b$ is the primitive lattice parameter and $m = 0,1,2,...$).We are 
considering the case where $(R..B..R..B...) \rightarrow (R...R...R...R....)$.

We have combined the colour labels (``up'' and ``down'') and the
position labels ($k$ and $k+(2m+1)b$) into a single column
vector. The column vector on the right hand side signifies that
there is an ``up'' spin $(R)$ at the position $k$ and a ``down''
spin $(B)$ at the position $k+(2m+1)b$ where $(2m+1)b$, $(m =
0,1,2,..$ etc.) is the lattice spacing. The column vector on the
left hand side has the corresponding meaning. The ordering we
have used in the initial configuration is (R, B). It can be
readily verified that the solution to the $V$ matrix is
$$V = \left(\begin{array}{cccc}
1 & 0 & 0 & 0\\
1 & 0 & 0 & 0\\
0 & 0 & \alpha - \beta/(2m+1)b & \beta/(2m+1)b\\
0 & 0 & - \beta/(2m+1)b & \alpha + \beta/(2m+1)b
\end{array}\right) \eqno(2)$$
Now consider the case $R \Leftrightarrow B$ in equation (1). The
matrix $V$ remains the same. This is expected as this reflects a
symmetry of the system. The Van der Waerden transformation
should be insensitive to what is called ``up'' and what is
called ``down''.

The final configuration can also consist of all ``down'' spins
starting with the initial ordering as (R, B) i.e.
$$\left(\begin{array}{l}
B_{\alpha k + \beta}\\
B_{\alpha[k+(2m+1)b]+\beta}\\
\alpha k + \beta\\
\alpha[k+(2m+1)b]+\beta
\end{array}\right) = 
\left(\begin{array}{cccc}
V'_{11} & V'_{12} & V'_{13} & V'_{14}\\
V'_{21} & V'_{22} & V'_{23} & V'_{24}\\
V'_{31} & V'_{32} & V'_{33} & V'_{34}\\
V'_{41} & V'_{42} & V'_{43} & V'_{44}\end{array}\right)
\left(\begin{array}{l}
R_{k}\\
B_{k+(2m+1)b}\\
k\\
k+(2m+1)b \end{array}\right) \eqno(3)$$
and the solution to the $V'$ matrix is now
$$V' = \left(\begin{array}{cccc}
0 & 1 & 0 & 0\\
0 & 1 & 0 & 0\\
0 & 0 & \alpha - \beta/(2m+1)b & \beta/(2m+1)b\\
0 & 0 & - \beta/(2m+1)b & \alpha + \beta/(2m+1)b
\end{array}\right) \eqno(4)$$
As before the solution to $V'$ remains unchanged under $B
\Leftrightarrow R$ in equation (3) for the same reasons.

As the final configuration is monochromatic, it can be easily
verified that for the system we are considering with periodic
boundary conditions, $\alpha = 2$ and $\beta = (2m + 1)b$. In
fact, it can be shown that the latter part of the theorem can
also be realised viz. $\alpha$ and $\beta$ are bounded by a
function  that depends only on $X$ and $m$ $^{4)}$.

Let us now consider the energy of this classical spin
configuration. The energy of the initial configuration $E_{i}$
may be written formally as some average of some hamiltonian $H =
\sum_{ij(i \not= j)} s_{i} J_{ij} s_{j}$ with nearest neighbour interactions
$$E_{i} = \displaystyle\sum_{{m,n}_{m \not= n}} < m \vert H \vert
n > \eqno(5)$$
The suggestive notation used in (5) is for convenience. To
evaluate this we split the chain  into blocks :
Expand (5) in terms of these blocks. The block energies are all
identical. If $N$ be the number of blocks (5) becomes :
$$\frac{E_{i}}{N} = <u^{T}_{1} \vert H \vert u_{2} >=
J_{21}(2m+1)b) B_{k+(2m+1)b} R_{k+2(2m+1)b} \eqno(6a)$$
where
$$\vert u_{1} > = 
\left(\begin{array}{l}
R_{k}\\
B_{k+(2m+1)b}\\
k\\
k+(2m+1)b \end{array}\right),
\vert u_{2} > = 
\left(\begin{array}{l}
R_{k+2(2m+1)b}\\
B_{k+3(2m+1)b}\\
k + 2(2m+1)b\\
k + 3(2m+1)b
\end{array}\right) \eqno(6b)$$
$$H = \left(\begin{array}{cccc}
0 & 0 & 0 & 0\\
J_{21} & 0 & 0 & 0\\
0 & 0 & 0 & 0\\
0 & 0 & 0 & 0
\end{array}\right) \eqno(6c)$$
The subscripts in (6a) are only there to show that these are
nearest neighbour interactions.

In a similar manner it may be shown that the final energy is 
$$\frac{E_{f}}{N'} = <v^{T}_{1}\vert H'\vert v_{2}>
\phantom{.....................}$$
$$\phantom{\frac{E_{f}}{N'}...............................} = J'_{21} (\alpha(2m + 1)b)
R_{\alpha[k+(2m+1)b]+\beta} R_{\alpha[k+2(2m+1)b]+\beta} \eqno(7a)$$
$$\vert v_{1} >= \left(\begin{array}{l}
R_{\alpha k+\beta}\\
R_{\alpha[k+(2m+1)b]+\beta}\\
\alpha k + \beta\\
\alpha[k+(2m+1)b]+\beta\end{array}\right),
\vert v_{2} >= 
\left(\begin{array}{l}
R_{\alpha[k+2(2m+1)b]+\beta}\\
R_{\alpha[k+3(2m+1)b]+\beta}\\
\alpha[k + 2(2m+1)b]+\beta\\
\alpha[k + 3(2m+1)b]+\beta
\end{array}\right) \eqno(7b)$$
$$H' = \left(\begin{array}{cccc}
0 & 0 & 0 & 0\\
J'_{21} & 0 & 0 & 0\\
0 & 0 & 0 & 0\\
0 & 0 & 0 & 0
\end{array}\right) \eqno(7c)$$

Let us write $< v^{T}_{1}\vert = < u^{T}_{1}\vert V^{T}$ and
$\vert v_{2} >= V\vert u_{2} >$, then 
$$\frac{E_{f}}{N'} = < v^{T}_{1}\vert H' \vert v_{2} > = <
u^{T}_{1}\vert V^{T} H' V\vert u_{2} >$$
$$\phantom{\frac{E_{f}}{N'}........................} = < u^{T}_{1}\vert H_{d}'\vert u_{2}
> = J'_{21}(\alpha(2m + 1)b) R_{k} R_{k+2(2m+1)b} \eqno(8)$$
where $H'_{d}$ =  diag $(J'_{21}(\alpha(2m + 1)b), 0, 0, 0)$.

So the VDW matrices diagonalises the hamiltonian. This is an
expected result as the final configuration being ferromagnetic
means that in ``colour'' space the hamiltonian should be
diagonal. The first two entries of our column vectors refer to
the ``colour'' space whereas the last two entries refer to the
one dimensional lattice. Accordingly, the upper left block of
the VDW matrix operates in the ``colour'' space while the lower
right block operates on the lattice. Since the interaction is
only in ``colour'' space, diagonalisation means that interaction
is between objects with identical ``colour'' (i.e. spins). The
separation of the nearest neighbour spins in (7a) is
$\alpha(2m+1)b$ i.e. $2(2m+1)b$ as $\alpha = 2$. This is
identical to the separation in (8) and thus confirms the
consistency of our basic formalism.

The antiferromagnetic chain has been mapped onto itself except
that now there is an effective ferromagnetic ordering. No
dynamics has been involved and so the initial and final energies
are equal. As usual, in the calculation of energies we take the
value of the ``up'' spin to be $+1$ and that for the ``down''
spin to be $-1$. The lattice is infinite, so $N = N'$. Equating
(6a) with (7a) we get :
$$J'_{21}(\alpha(2m + 1)b) = - J_{21}((2m + 1)b)$$
i.e.
$$J'_{21}(2(2m + 1)b) = - J_{21}((2m + 1)b) \eqno(9)$$

In the one dimensional Ising model $^{2)}$ , the antiferromagnetic
Ising model in the thermodynamic limit (with nearest neighbour
interactions and periodic boundary conditions) and with
hamiltonian (energy)
$$E = J \sum_{i} s_{i} . s_{i+1} \eqno(10a)$$
has the same ground state energy as the effective ferromagnetic
model with only next-nearest-neighbour interactions (in terms of
original lattice) and with hamiltonian (energy)
$$E = - J' \sum_{i} s_{i} . s_{i+2} \hspace{.5in}with \hspace{.2in} J' = J \eqno(10b)$$

Note that the nearest neighbour ferromagnetic model obtained by
the ``staggering'' operation $^{3)}$ from the original
antiferromagnetic model has energy
$$E = - J" \sum_{i} s_{i} . s_{i+1}
\hspace{.5in} with \hspace{.2in} J" = J \eqno(11)$$
$(J \rightarrow - J, s_{i} \rightarrow (-1)^{i+1} s_{i},  E$ is
invariant)

In the VDW case there is no flip in spin space but a translation
in coordinate space to a point which has a flipped spin and so
the resulting effective ferromagnetic model has a pseudo next
nearest neighbour interactions (with respect to
original lattice). However, in the staggered case there is no
translation in coordinate space while there is a flip in spin
space and so the resulting ferromagnetic model still has nearest
neighbour interactions.

In the VDW mapping, however, $J' (\alpha b) = J(b)$ for $\alpha
= 2, \beta = 1$. So the VDW mapping is similar in spirit to the
model embodied in equations (10). Note that the equality of
initial and final energies is an input valid only in the
thermodynamic limit. So the VDW transformation is like a
symmetry operation in this limit.

The VDW theorem has enabled us to relate the coupling constants
in a simple way. Although the problem considered here was rather
simple, {\it we have achieved some flavour of the usual coarse
graining (block) average-the precursor to the renormalisation
group. The equality of the energies are similar to the demand
of the invariance of the partition function. Also note that in
our scheme the inverse of the matrices $V$ and $V'$ do not
exist}. Under the VDW transformation, translational invariance is
preserved - $\beta$ is essentially a shift and hence does not in
anyway appear in the energies or coupling constants.

However, all the transformations are on the vectors and not on
the hamiltonian. Also the product of successive VDW
transformations do not make sense. This is because after the
very first transformation monochromaticity is obtained and so
the premises of the theorem are fulfilled. Further
transformations are meaningless.

It can be shown that {\it at the level of the matrices} for $\alpha =
2$ and $\beta = (2m + 1)b$
$$V^{2}((2m + 1)b) = 2 V((2m + 1)b/2)$$

{\it In this sense the VDW transformations seem to mimic the group
property if we consider the elements of the group as the set of
all VDW transformations.The above expression depicts the functional relation
for the representation of VDW theorem in the product space of colour and
the one dimensional lattice.}

One can also show the following: Considering a two dimensional spin system as an 
infinite collection of spin chains , the topological property of 
any interface $^{5)}$ is unchanged after a VDW transformation $^{4)}$.

All the above results have  been extended to
the case with three colours$^{4)}$ and hence the entire gamut of ternary alloys
can now be looked at in a new light.In fact , the exact solutions obtained 
$^{4)}$ opens new vistas in the study of disordered lattices.
Moreover,the quantum version of our work is presently under consideration.

The authors would like to thank Sukumar Das Adhikary for
introducing them to VDW Theorem.

\end{document}